\documentclass[12pt]{iopart}

\usepackage[utf8]{inputenc}
\usepackage[T1]{fontenc}
\usepackage[english]{babel}
\usepackage{iopams}
\usepackage{graphicx}
\newcommand{\onefig}[1]{\centering{
    \includegraphics[width=0.8\columnwidth]{#1}}}

\newcommand{\myii}{\mathrm{i}}

\newcommand{\matr}[1]{\boldsymbol{\mathrm{#1}}}

\renewcommand{\Re}{\mathop{\mathrm{Re}}\nolimits}
\renewcommand{\Im}{\mathop{\mathrm{Im}}\nolimits}

\newcommand{\mean}[1]{\langle #1\rangle}

\newcommand{\erfc}{\mathop{\mathrm{erfc}}\nolimits}

\newcommand{\clvecfwd}{\vec{\gamma}}

\newcommand{\spanspc}{\mathop{\mathrm{span}}}
\newcommand{\spaceA}{\mathcal{A}}
\newcommand{\spaceB}{\mathcal{B}}
\newcommand{\mycolon}{\,\mathrm{:}\,}

\begin{document}

\title[Violation of hyperbolicity via UDV]{Violation of hyperbolicity via
  unstable dimension variability in a chain with local hyperbolic
  chaotic attractors}

\author{Pavel V Kuptsov}%
\address{Department of Instrumentation Engineering, Saratov State
  Technical University, Politekhnicheskaya 77, Saratov 410054,
  Russia}%
\ead{p.kuptsov@rambler.ru}

\begin{abstract}
  We consider a chain of oscillators with hyperbolic chaos coupled via
  diffusion. When the coupling is strong the chain is synchronized and
  demonstrates hyperbolic chaos so that there is one positive Lyapunov
  exponent. With the decay of the coupling the second and the third
  Lyapunov exponents approach zero simultaneously. The second one
  becomes positive, while the third one remains close to zero. Its
  finite-time numerical approximation fluctuates changing the sign
  within a wide range of the coupling parameter. These fluctuations
  arise due to the unstable dimension variability which is known to be
  the source for non-hyperbolicity. We provide a detailed study of
  this transition using the methods of Lyapunov analysis.
\end{abstract}


\vspace{2pc}


\submitto{\JPA}


\section*{Introduction}

Structural stability is the fundamental property of dynamical systems
that implies that qualitative behavior of a system is preserved under
perturbations. Structurally stable systems are really significant for
theoretical and numerical researches, and especially for practical
applications~\cite{Andronov, Shilnikov}. Simple examples of
structurally stable dynamics provide fixed points and limit
cycles. The chaotic dynamics is not structurally stable in general;
however, a special class of systems with uniform hyperbolicity
possesses this property. Mathematical examples of hyperbolic chaotic
systems are known since 1960s~\cite{Anosov,Smale,Williams,Plykin}, but
physically realizable systems with attractors of this type were
discovered sufficiently recently. As reported
in~\cite{Hyp,HypExper,KuzBook,KuzUfn}, simple systems of coupled
oscillators exciting alternately possess uniformly hyperbolic
attractors of Smale-Williams type. Chaos in these systems is related
to dynamics of the phases of oscillators in the successive stages of
activity governed by a Bernoulli-type expanding circle map.

An attractor is said to be uniformly hyperbolic if the tangent space
associated with any of its point can be split into a direct sum
decomposition of uniformly expanding and contracting subspaces and
this splitting is invariant under the tangent
mapping~\cite{KatHass}. Tangent vectors from the expanding
(contracting) subspace always grow (decay) in time, i.e.,
corresponding finite-time Lyapunov exponents (FTLEs) are strictly
positive (negative). In particular, it means that angles between the
expanding and contracting tangent vectors are well separated from
zero. The violation of the uniform hyperbolicity can occur due to the
emerging of homoclinic tangencies, i.e., tangential intersections of
stable and unstable manifolds of the attractor. Trajectories with the
\emph{exact} tangencies are rather untypical in a sense that we almost
never hit them choosing random initial conditions. But if they exist,
any trajectory can pass infinitely close to them so that the angles
between expanding and contracting tangent vectors can be arbitrary
small.

A stronger form of non-hyperbolicity is unstable dimension variability
(UDV), which is characterized by the coexistence, in the chaotic
attractor, of invariant periodic or chaotic orbits with a different
number of unstable directions~\cite{UDVMain,PereiraPinto}. Since
trajectories of the system can pass close to these orbits, the
dimension of their unstable and stable manifolds varies. Thus the
attractor with UDV does not admit the invariant splitting of the
tangent space into strictly expanding and strictly contracting
subspaces, so that the conditions of the uniform hyperbolicity fail.

Due to the variations of the unstable manifold dimension, the closest
to zero FTLE fluctuates changing its
sign~\cite{SauerGrebogi97,UDVMain,PereiraPinto}. Its distribution
spreads both in positive and in negative semi-planes, and the equal
probabilities for positive and negative values of the FTLE indicate
the strongest UDV~\cite{Sauer02}.

Floating point computations employed in modeling of a dynamics are
always subjected to errors caused by finite-precision computer
arithmetic. If the dynamics is chaotic, numerical trajectory diverges
from the modeled true trajectory. However, if the system holds the
shadowing property, there exists a true chaotic trajectory which stays
uniformly close to the numerical one for a certain time interval. This
shadowing trajectory may not correspond to the initial conditions
which was used to compute the numerical trajectory, however its very
existence indicates that the computed statistical results are
valid. For hyperbolic chaotic systems the time of shadowing is
unlimited~\cite{Ano67,Bowen75}, and most of chaotic systems with
homoclinic tangencies also have a reasonably long shadowing
time~\cite{GrebHam90}. For systems with UDV the shadowing time can be
small which is a serious obstacle for computer
modelings~\cite{SauerGrebogi97,UDVMain,PerCam08}.
In~\cite{SauerGrebogi97} the estimates for shadowing time and distance
are found to be functions of computer round-off error and variance of
the sing-changing FTLE. In~\cite{Sauer02} the scaling law is suggested
and the characteristic hyperbolicity exponent is introduced that
describe the growth of errors of trajectory averages for systems with
the sign-changing FTLE. Altogether, real systems with UDV present many
challenges for theoretical and experimental
investigation~\cite{Romeiras,PerCam08}.

One of the typical situation where UDV occurs is the loss of
synchronization of identical coupled chaotic
systems~\cite{PereiraPinto,UDVCplChaSyst,UDVSycnChaSyst}. As argued
in~\cite{BarErn2000} there are two mechanisms for UDV in this case.
While the coupling is sufficiently strong so that the overall
synchronization attractor is stable, its embedded periodic orbits can
lose the transverse stability. Since for different orbits it can
happen at different coupling strengths, it constitutes the UDV
mechanism of the first type. In this case a trajectory, started
outside but close to the synchronization attractor, quietly approaches
it for some time, but then it can pass near a transversally unstable
orbit that results in chaotic burst. This behavior, referred to as
on-off intermittency~\cite{OnOff,PikSyn}, is the one of observable
manifestations of UDV. The embedded unstable periodic orbits undergoes
bifurcations as the coupling decays so that new saddle and repelling
orbits appear outside of the synchronization manifold. The set of new
orbits is called the emergent set~\cite{BarErn2000}. Beyond the
blowout bifurcation~\cite{Blowout} where the synchronization attractor
becomes transversally repelling, the most of the emergent set is
incorporated into the new non-synchronous attractor. Since the orbits
from this set can have different dimensions of their unstable
manifolds, the UDV occurs, and this is the second mechanism.

Recently the interest to high-dimensional or spatially extended
hyperbolic systems has been renewed after the discovery of the
numerical algorithms for covariant Lyapunov vectors (CLVs)
(see~\cite{GinCLV,WolfCLV} for the original papers,
and~\cite{CLV2011,GinChatLivPol12} for reviews). One way to obtain
such a system is to build it by coupling together low-dimensional
hyperbolic chaotic subsystems. Examples are coupled map lattices with
hyperbolic chaos considered in~\cite{Sinai88,Bunimovich93} or the
diffusive medium with local uniform hyperbolicity suggested
in~\cite{HyperSpace08}. Less straightforward approach is to identify
or create in an extended system the mechanism of spatial modes
interactions resulting in hyperbolic chaos. This approach is
implemented in~\cite{HypTur2012} for Turing patterns.

Previously, in paper~\cite{HyperSpace08} we considered the diffusive
medium with local hyperbolic attractor and with no-flux boundary
conditions. When spatial coupling is strong, the medium is
synchronized and demonstrates uniformly hyperbolic chaos. This regime
is characterized by the single positive Lyapunov exponent.  Decreasing
the coupling results in the desynchronization accompanied by the
emerging of the second positive Lyapunov exponent. However, the
hyperbolicity survives, so that the system demonstrates a kind of
hyperbolic hyper-chaos. The violation of the hyperbolicity occurs only
when the third Lyapunov exponent becomes positive.

In this paper we study a system which is very similar to that one
presented in~\cite{HyperSpace08}. Namely, this is a chain of
oscillators with hyperbolic chaos of Smale-Williams type coupled
diffusively. The major difference is that the boundary conditions are
now periodic. This results in the degeneration of the Lyapunov
spectrum so that the second and the third exponents coincide in the
synchronous regime. Decreasing the coupling results in their
simultaneous approaching to zero where the synchronization is broken.
Above the desynchronization threshold the second Lyapunov exponent
grows while the third one stays near zero. Numerical, i.e., finite
time, approximation of this Lyapunov exponent largely fluctuates
frequently changing its sign, and this happens for a wide range of
coupling strengths. This behaviour is a result of UDV. Subsequent
decay of the coupling strength results in the growth of the third
Lyapunov exponent and thus to the ceasing of UDV.

The desynchronization of a chain of chaotic elements via UDV is
studied in~\cite{ViaGreb03} for the lattice of coupled logistic
maps. The coupling is varied from the nearest neighborhood to the
global coupling. The basic phenomenology related to this system is
revealed and discussed, such as UDV, on-off intermittency, FTLE
fluctuations and shadowing. In our investigation below, instead of
maps we consider a chain of oscillators that are uniformly hyperbolic
and admit physical implementation~\cite{HypExper}. Moreover, we
concentrate on UDV effect occurring beyond the blowout bifurcation due
to unstable invariant orbits coexisting with the non-synchronous
attractor.

The paper is organized as follows. In \sref{sec:local} the local
oscillators with hyperbolic chaos are discussed, \sref{sec:chain}
introduces the system under consideration, and in \sref{sec:main} we
consider the details of the transition to the non-hyperbolic
chaos. Namely, we show how the distance to the synchronization
manifolds grows as the coupling strength decays, \sref{sec:dist};
analyze Lyapunov exponents as well as FTLEs, sections~\ref{sec:le} and
\ref{sec:ftle}, respectively; then we compute angles between tangent
subspaces, \sref{sec:hyp}; finally, discuss the structure of CLVs,
\sref{sec:clv}. \Sref{sec:sum} summarizes the results.

\section{Local oscillators with hyperbolic chaos}\label{sec:local}

The building block for our model is a set of amplitude equations for
the system of two coupled van der Pol oscillators with alternating
excitation that is known to demonstrate uniformly hyperbolic
chaos~\cite{Hyp,KuzUfn,KuzBook,Wilczak10}:
\begin{equation}
  \label{eq:ode}
  \eqalign{%
    \dot{a}= Aa\cos(2\pi t/T)-|a|^2 a-\myii\epsilon\,b, \\
    \dot{b}=-Ab\cos(2\pi t/T)-|b|^2 b-\myii\epsilon\,a^2.}
\end{equation}
Here $a$ and $b$ are complex dynamical variables, $A$ controls the
magnitude of the excitation, $\epsilon$ is the coupling parameter and
$T$ is the period of excitation. The subsystems are excited in the
alternating manner. The period of the excitation $T$ is assumed to be
large with respect to the duration of transient processes in the
subsystems controlled by $A$. Suppose at some instant the first
oscillator is excited and its amplitude $|a|$ is high. Hence, the
second one is suppressed so that its amplitude $|b|$ is small. The
coefficients in~\eref{eq:ode} are real except for the coupling. It
means that the phases can be changed only as a result of the
interaction between the subsystems. But when $a$ is excited, $|b|$ is
small, and its action on $a$ is negligible. Thus, the phase of $a$
remains constant during the excitation stage. The backward influence
from $a$ to $b$ is high, and the coupling term is proportional to
$a^2$. It means that after the interval $T/2$ the oscillator $b$ at
the onset of its own excitation inherits a doubled phase of $a$ (the
phase also gets a shift $-\pi/2$ because of the imaginary unit at the
coupling term). Now the roles of the subsystems are exchanged. The
phase of $b$ remains constant when this subsystem is excited, and at
the end, after the other lapse $T/2$, the phase is returned back to
$a$ through a linear coupling term (also with the shift $-\pi/2$). As
a result, the first oscillator $a$ doubles its phase during the
period~$T$.

The above discussion allows us to write down a stroboscopic map for a
series of phases $\phi_n = \arg a(nT)$ that are measured after each
period~$T$: $\phi_{n+1}=2\phi_n - \pi \mathop{\mathrm{mod}} 2\pi$.
This map demonstrates uniform hyperbolic chaos: the rate of
exponential divergence of two close trajectories is identical in each
point of the phase space and equal to $\ln 2$. Since the described
mechanism is the solely responsible for the chaos in the
system~\eref{eq:ode}, one can expect that the system itself
demonstrates hyperbolic chaos. The detailed analysis presented
in~\cite{KuzBook,KuzArt} confirms that the stroboscopic map, whose
variables coincide with $a(t)$ and $b(t)$ at $t=t_n=nT$ is indeed
hyperbolic. In particular, the largest Lyapunov exponent of this map
is very close to $\Lambda_1=\ln 2$.

\section{The model system}\label{sec:chain}

First, using the oscillators~\eref{eq:ode} as local elements, we
construct an active diffusion medium studied in~\cite{HyperSpace08}:
\begin{equation}
  \label{eq:pde}
  \eqalign{%
    \partial_t a= A\cos(2\pi t/T)a-|a|^2 a-
    \myii\epsilon\,b+\partial_x^2 a,\\
    \partial_t b=-A\cos(2\pi t/T)b-|b|^2 b-
    \myii\epsilon\,a^2+\partial_x^2 b.}
\end{equation}
Since we are going to solve these equations numerically, the
discretization of spatial derivatives can be taken into account
explicitly. Thus we obtain a chain of $N$ hyperbolic oscillators with
the diffusive coupling:
\begin{equation}
  \label{eq:chain}
  \eqalign{%
    \dot{a}_i= A\cos(2\pi t/T)a_i-|a_i|^2 a_i-
    \myii\epsilon\,b_i+\kappa(a_i)/h^2,\\
    \dot{b}_i=-A\cos(2\pi t/T)b_i-|b_i|^2 b_i-
    \myii\epsilon\,a_i^2+\kappa(b_i)/h^2.}
\end{equation}
Here $\kappa(z_i)=z_{i-1}-2z_i+z_{i+1}$, $i=0,\ldots N-1$ and $h$ is
the step of discretization. The length of the system is $S=Nh$, and
the diffusion coefficients are rescaled to units. The spatial coupling
in this chain is controlled by varying $S$. While for the original
continuous model~\eref{eq:pde} this is an unambiguous manipulation,
dealing with numerical simulations we have two main strategies which,
in principle, can give different results. The first one is to vary $N$
while keeping constant $h$. In this way the thermodynamic limit can be
investigated where the number of degrees of freedom is infinite. The
second approach is to vary $h$ at constant $N$ and thus to investigate
the continuous limit at $h\to 0$. One more reasonable approach is to
consider the continuous limit at constant $S$: increasing of $N$ and
proportional decreasing of $h$. Below we shall vary $h$ at constant
$N$.

Thus, parameter $h$ plays the role of a reciprocal coupling strength,
i.e., high values of $h$ correspond to weak coupling and vice
versa. The lowest reasonable values of $h$ are those for which the
chain demonstrates full chaotic synchronization. The highest ones are
attained when the oscillators are effectively uncoupled. Each partial
oscillator has one positive Lyapunov exponent. As we shall see below,
the number of positive Lyapunov exponents for the whole chain grows
with $h$. Thus, the chain is effectively uncoupled when the total
number of positive exponent is equal to $N$.

In~\cite{HyperSpace08} equations~\eref{eq:pde} was studied with the
no-flux boundary conditions. In the present work we consider the
chain~\eref{eq:chain} with the periodic boundary conditions.

Since the uniform hyperbolicity is proven only for the stroboscopic
map corresponding to~\eref{eq:ode}, in what follows we shall represent
the results of simulations of~\eref{eq:chain} in terms of the
stroboscopic map whose state variables coincide with the chain
variables $a_i(t)$ and $b_i(t)$ at $t=t_n=nT$. In fact, it merely
means that the Lyapunov exponents $\lambda_i$ are computed first for
the continuous time system~\eref{eq:chain}, and then they are
multiplied by the time step $T$, i.e., $\Lambda_i=T\lambda_i$. In the
same way FTLEs of~\eref{eq:chain}, computed for time intervals $nT$,
are related to FTLEs $L_i(n)$ of the stroboscopic map. Finally, the
Lyapunov vectors for the stroboscopic map as well as their angles are
obtained for the continuous time system at $t=nT$.

\section{Transition from hyperbolic to non-hyperbolic chaos via
  unstable dimension variability}\label{sec:main}

\subsection{The distance to the synchronization manifold}\label{sec:dist}

Small $h$ produces strong coupling. In this case the oscillators
of~\eref{eq:chain} are synchronized, and the stroboscopic map for the
chain demonstrates hyperbolic chaos with the single positive Lyapunov
exponent $\Lambda_1\approx \ln 2$. The attractor in the synchronous
regime will be referred to as synchronization attractor, and the
manifold containing this attractor $a_i=a_j$, $b_i=b_j$, where $0\leq
i,j\leq N-1$, will be referred to as the synchronization manifold.

Increasing $h$ we expect to observe the destroying of the synchronous
regime. To characterize the deviation of the system from the
synchronization manifold it is convenient to use new variables
measured along directions transverse to this manifold, that are
introduced in~\cite{KupKuz04}. (Another formulas for transverse
directions can be defined, see for example~\cite{TransvDynam}.)  There
are $N-1$ such variables that can be computed as
\begin{equation}
  \label{eq:transv_coord}
  a'_i(t)=a_i(t)-\frac{1}{i}\sum_{k=0}^{i-1}a_k(t),
\end{equation}
where $i=1,\ldots N-1$. In the same way they are introduced for
$b_i(t)$\footnote{To obtain a complete set of new variables, one have
  to add here the longitudinal variables $a'_0$ and $b'_0$ computed as
  averaged $a_i$ and $b_i$, respectively.}.  Now given the transverse
coordinates, we can find the distance from the synchronization
manifold as
\begin{equation}
  \label{eq:transv_dist}
  \rho(t)=\sqrt{\sum_{i=1}^{N-1}|a'_i(t)|^2+|b'_i(t)|^2}
\end{equation}
To compute the distance corresponding to the stroboscopic map we have
to take into account variables $a_i(t)$ and $b_i(t)$ only at $t=nT$.

When $h\leq 1.04$ the distance $\rho(nT)$ vanishes, i.e. the chain is
synchronized. We note that the full synchronization of $N$ chaotic
elements with the local coupling is possible only when $N$ is finite
and below some threshold.  Paper~\cite{ViaGreb03} shows that given $N$
one can obtain the synchronization increasing the coupling
range. In~\cite{KupKuz04} we demonstrate that the number of elements
that can be synchronized depends on the largest Lyapunov exponent of
the partial element: the higher is the exponent, the smaller is $N$.

\begin{figure}
  \onefig{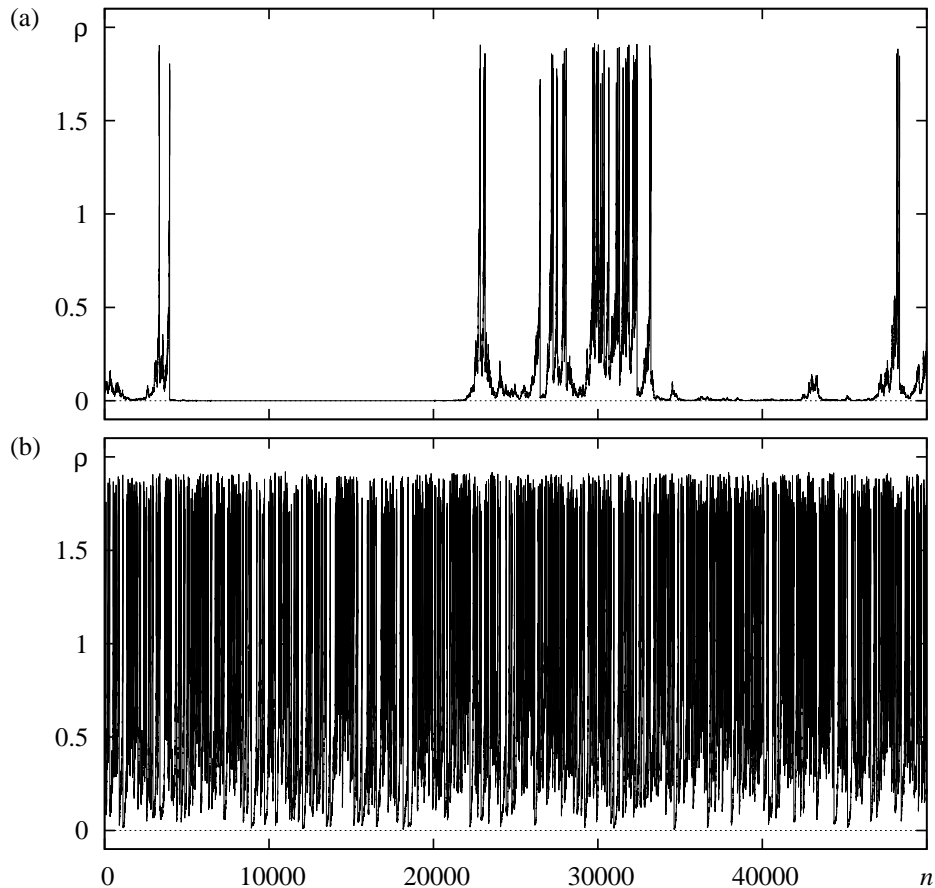}
  \caption{\label{fig:rhotime}The distance $\rho(nT)$ to the
    synchronization manifold vs. time step $n$ for (a) $h=1.05$, and
    (b) $h=1.06$. $N=16$, $A=3$, $T=5$, $\epsilon=0.05$.}
\end{figure}

Figure~\ref{fig:rhotime} shows the time dependence of $\rho(nT)$ above
the desynchronization threshold. In Fig.~\ref{fig:rhotime}(a) we
observe alternating quiescent and bursting phases. This regime is
called on-off intermittency~\cite{OnOff,OnOffNetw}. It is typical for
coupled systems losing chaotic synchronizations. Unstable periodic
orbits embedded into the synchronous attractor becomes transversally
unstable before the attractor as a whole do so. As a result a
trajectory not exactly belonging to the attractor may approach to it
for sufficiently long time, but the passings close to points of the
unstable orbits result in the burstings. Since the embedded periodic
orbits and the attractor as a whole have different dimensions of
unstable and stable manifolds, the UDV takes place, which, in
particular, results in the violation of the uniform
hyperbolicity~\cite{UDVMain,PereiraPinto}. According to the notation
of~\cite{BarErn2000}, this is the first mechanism of UDV.

Figure~\ref{fig:rhotime}(b) is plotted for a larger value of $h$. The
intermittency is absent, and irregular oscillations are observed
instead. It means that now the synchronous attractor as well as the
most of the unstable periodic orbits are transversally unstable so
that trajectories never stay close to the synchronization manifold for
a long time. Notice that $h$ in this panel is just a little bit higher
then in the panel (a). It means that the transverse destabilization of
embedded periodic orbits takes place for our system within a very
narrow range of the coupling parameter. Previously we already observed
the very narrow range of on-off intermittency and riddling for a
system of two coupled oscillators~\eref{eq:ode}, see~\cite{KupKuz06}.
Also the narrow range of the on-off intermittency regime is reported
in~\cite{ViaGreb03} for the lattice of coupled logistic maps.

Averaging $\rho(t)$ over a large number of trajectory points results
in the characteristic value $\mean\rho$ indicating how far the system
is off the synchronization manifold. Figure~\ref{fig:lyap}(a) shows
the growth of $\mean \rho$ as $h$ grows. The transition to
non-synchronous regime occurs in the vicinity of $h\approx 1.04$.

\subsection{Lyapunov exponents and the manifestation of UDV}\label{sec:le}

\begin{figure}
  \onefig{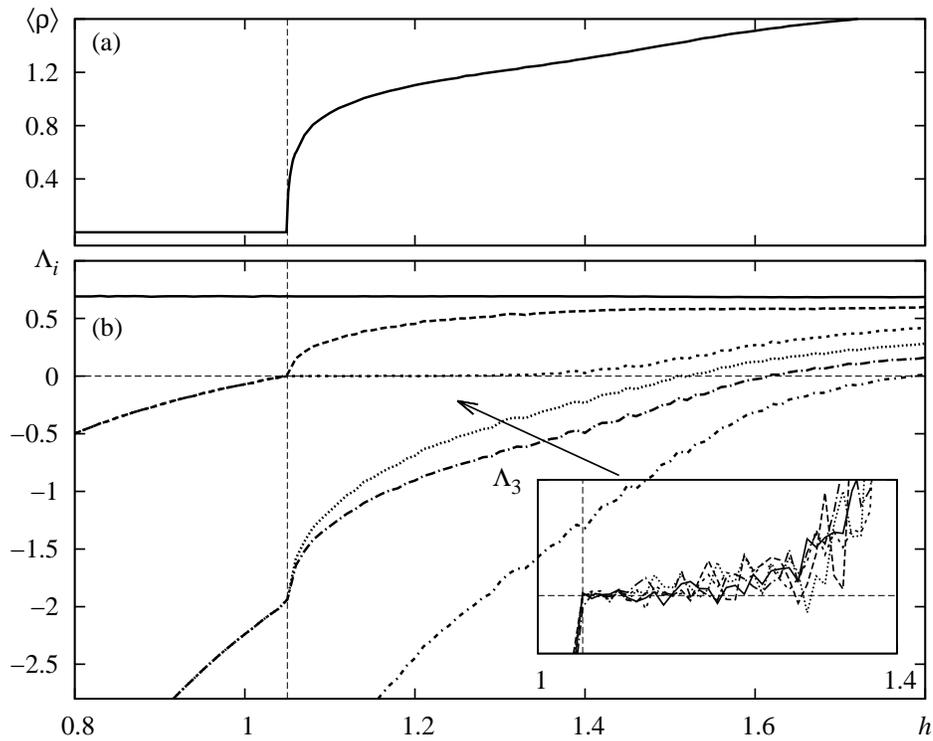}
  \caption{\label{fig:lyap}(a) The average distance $\mean\rho$ form
    the synchronization manifold and (b) the first six Lyapunov exponents
    vs. the reciprocal coupling strength $h$. The inset in the
    panel (b) shows $\Lambda_3$ in the larger scale. $N=16$, $A=3$,
    $T=5$, $\epsilon=0.05$.}
\end{figure}

Now we turn to the Lyapunov exponents. Figure~\ref{fig:lyap}(b)
demonstrates the first six Lyapunov exponents versus the coupling
parameter $h$. First of all notice that the largest one remains almost
constant, $\Lambda_1\approx \ln 2$. In the synchronous regime all
other exponents are negative. Due to the symmetry imposed by the
periodic boundary conditions, the second and the third Lyapunov
exponents coincide, $\Lambda_2=\Lambda_3$. When $h$ grows these
exponents approach zero simultaneously. Then the system leaves the
synchronization manifold that results in the breakdown of the
degeneracy. The second exponent becomes positive, while the absolute
value of the third one remains small.

\begin{figure}
  \onefig{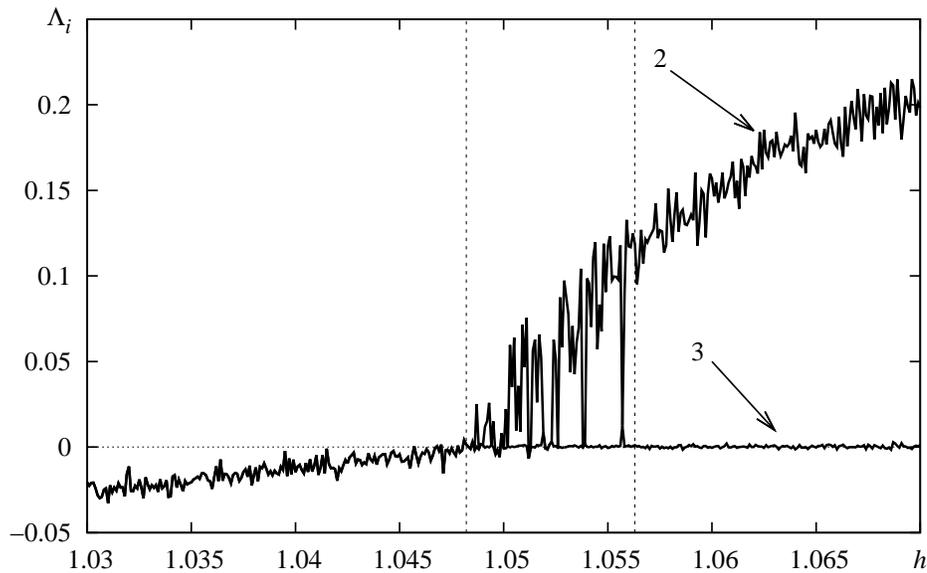}
  \caption{\label{fig:lyapzoom}The second and the third Lyapunov
    exponents vs. $h$ in the area of the desynchronization. Dashed
    vertical lines mark the range with on-off intermittency.}
\end{figure}

Figure~\ref{fig:lyapzoom} shows the behavior of the second and the
third Lyapunov exponents in a close vicinity of the desynchronization
onset. One can see the fluctuations of the numerical approximation of
$\Lambda_2$, as marked by the dashed vertical lines. The sign-changes
are very rare, but nevertheless occur. Moreover, here the on-off
intermittency is observed, as seen in Fig.~\ref{fig:rhotime}(a). This
agrees with our above discussion of the UDV arising in this area due
to the loss of the transverse stability of the unstable periodic
orbits embedded into the synchronization attractor.

To the right of the marked area in Fig.~\ref{fig:lyapzoom} the
numerical approximation of $\Lambda_2$ does not fluctuate anymore. The
inset in Fig.~\ref{fig:lyap}(b) shows however that $\Lambda_3$ changes
its sign. Inspecting the results of five independent computations of
$\Lambda_3(h)$ one can see that the numerical approximation of
$\Lambda_3$ often passes up and below the zero axis as $h$ grows. The
fluctuations occur within a wide range of $h$ and disappear at
$h\approx 1.4$ where the exponent leaves the vicinity of zero. These
fluctuations indicate the presence of UDV arising here via the second
mechanism due to the emergent sets appearing outside the
synchronization attractor~\cite{BarErn2000}.

Further growth of $h$, i.e., the decay of the coupling strength,
results in the vanish of UDV. The third Lyapunov exponent leaves the
vicinity of zero in a smooth way so that the probability of its sign
change vanishes. Then, as the coupling strength decays further, the
Lyapunov exponents becomes positive one by one, which is typical for
spatio-temporal chaotic systems.

\subsection{Fluctuations of finite-time Lyapunov
  exponents}\label{sec:ftle}

Fluctuations of $\Lambda_3$ observed in Fig.~\ref{fig:lyap}(b) are
obviously explained by the fact that numerically we can obtain only a
finite-time approximation of Lyapunov exponents. Thus, it is natural
to take these fluctuations into account explicitly, and discuss the
FTLEs $L_i(n)$, such that
\begin{equation}
  \Lambda_i=\lim_{n\to \infty} L_i(n).
\end{equation}
We recall that in fact the FTLEs will be computed for the continuous
times system~\eref{eq:chain} for time interval $nT$.

\begin{figure}
  \onefig{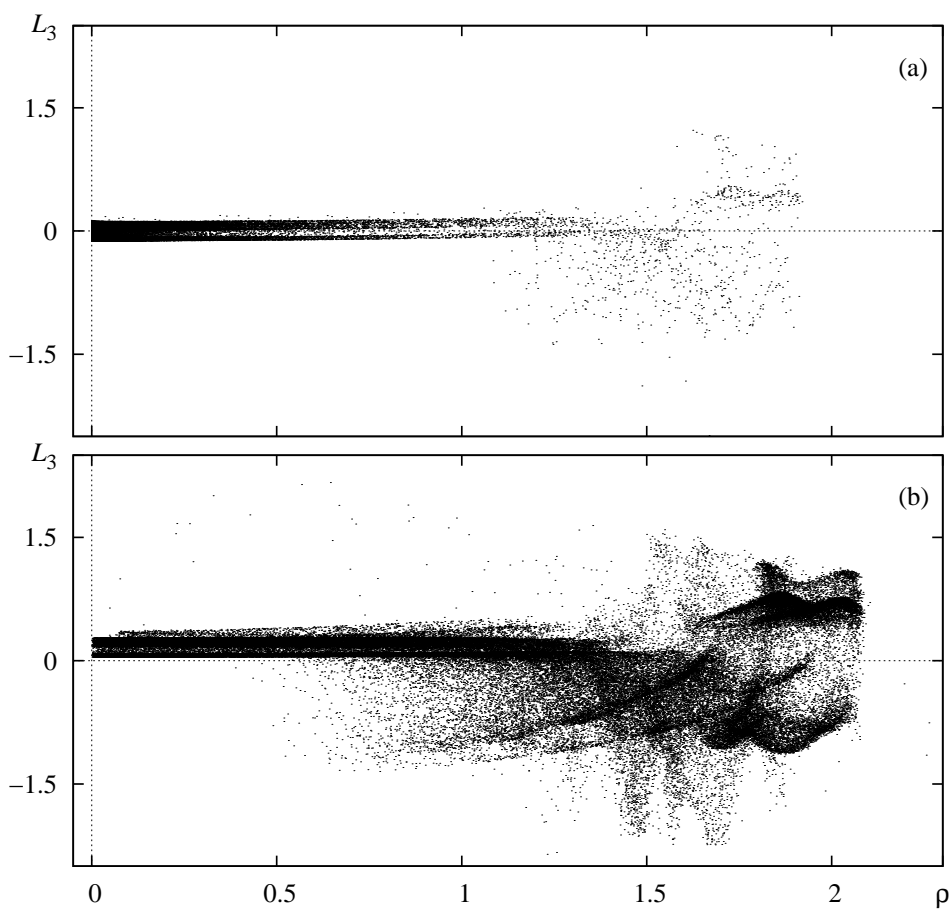}
  \caption{\label{fig:rholam} The third FTLE $L_3(1)$ vs.
    transverse distance to the synchronization manifold $\rho$
    computed for $10^5$ trajectory points. (a) $h=1.05$, (b) $h=1.2$.}
\end{figure}

Figure~\ref{fig:rholam} shows the plot of $L_3(1)$ against the
distance $\rho$ to the synchronization manifold for $10^5$ trajectory
points. The panel (a) corresponds to the on-off intermittency case,
shown in Fig.~\ref{fig:rhotime}(a). The bulk of points is located near
the synchronization manifold and $L_3(1)$ fluctuates there, changing
its sign. This again confirms that the first mechanism of UDV
dominates in this case, i.e., the UDV appears due to the periodic
orbits embedded into the synchronization manifold. Nevertheless, there
are rarer fluctuations of high amplitude at high $\rho$. They appear
due to the second mechanism, i.e., due to the emergent set of orbits
outside the synchronization manifold. The panel (b) also corresponds
to the UDV case, but at higher $h$. Now $L_3(1)$ at small $\rho$ can
only approach zero axis from above without crossing it. It means that
all embedded periodic orbits are transversally unstable, and the first
UDV mechanism does not work. The UDV occurs solely due to the second
mechanism.

\begin{figure}
  \onefig{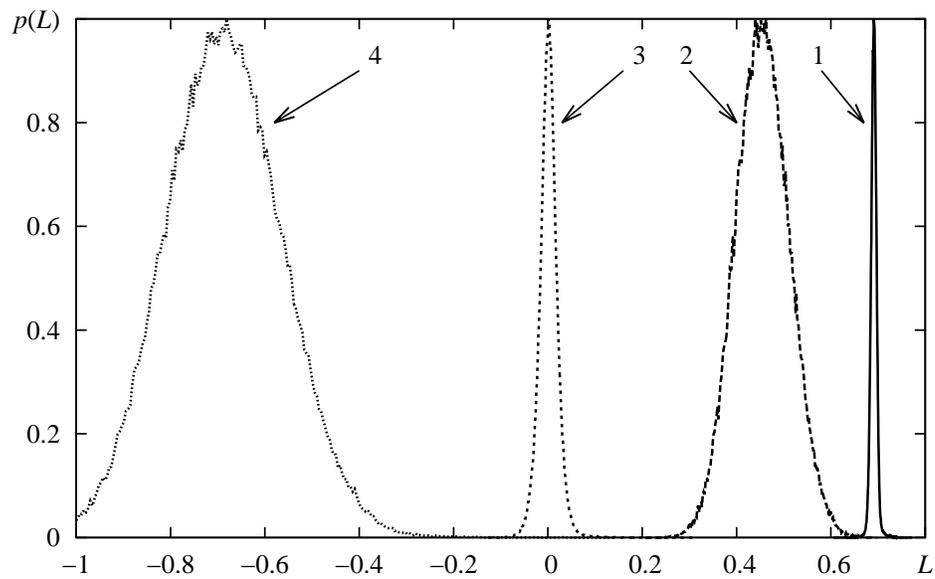}
  \caption{\label{fig:lldst} Distributions of the first four FTLEs
    $L_i(128)$ at $h=1.2$.}
\end{figure}

Figure~\ref{fig:lldst} shows the distributions of the first four FTLEs
$L_i(128)$ for the case represented in \ref{fig:rholam}(b). The
distributions have a Gaussian form: there are a well defined maxima
and rapidly decaying tails. Hence the infinite-time Lyapunov exponents
has definite values. The third exponent can take both positive and
negative values with almost equal probabilities. The detailed
inspection of the distribution reveals that the maximum of the curve
is located at $0.00168$ which can be taken as an approximate value of
$\Lambda_3$.

To characterize statistical properties of FTLEs, we use the approach
of~\cite{StatMechLyap11}. Employing the Gaussian approximation for
distributions, we compute covariances of all pairs $nL_i(n)$ and
$nL_j(n)$ as functions of $n$, then estimate their linear
slopes. These slopes are the diffusion coefficients describing the
growth rates of (co)variances of $nL_i(n)$. We use them to construct a
symmetric diffusion matrix $\matr D$. In fact, below we shall use only
the diagonal and sub-diagonal elements of this matrix. The variance of
the distribution of $L_i(n)$ is $D_{ii}/n$, where $D_{ii}$ is the
$i$-th diagonal element of the matrix. For the distributions shown in
Fig.~\ref{fig:lldst} the diffusion coefficients are $D_{11}=0.00461$,
$D_{22}=0.430$, $D_{33}=0.0455$, $D_{44}=2.05$.

Let $\ell$ be an index of the smallest positive infinite-time Lyapunov
exponent, and $\ell+1$ corresponds to the largest negative
one. Consider the probability of fluctuation of unstable direction
$P[L_\ell(n)\leq 0 \lor L_{\ell+1}(n)\geq 0]$. Since Gaussian
bivariate cumulative distribution function has a sufficiently
complicated closed form expression~\cite{Bivariate}, we shall estimate
its lower $P_0$ and upper $P_1$ bounds instead as follows:
\begin{equation}
  \eqalign{%
    P_0<P[L_\ell(n)\leq 0 \lor L_{\ell+1}(n)\geq 0]<P_1,\\    
  P_0=\max\{P[L_\ell(n)\leq 0], P[L_{\ell+1}(n)\geq 0]\},\\
  P_1=P[L_\ell(n)\leq 0]+P[L_{\ell+1}(n)\geq 0],}
\end{equation}
\begin{equation}
  \label{eq:prob_ftle}
  \eqalign{%
    P[L_i(n)\leq 0]=\erfc\left(\sqrt{n}H_i\right)/2, \\
    P[L_i(n)\geq 0]=\erfc\left(-\sqrt{n}H_i\right)/2,}
\end{equation}
where $H_i=\Lambda_i/\sqrt{2D_{ii}}$, and ``$\erfc$'' stands for the
complementary error function. The coefficient $H_i$ determines the
rate of probability convergence to the asymptotic value as $n$
grows. $|H_i|$ computed for the closest to zero FTLE is related to
hyperbolicity exponent $\tilde{h}$, introduced in~\cite{Sauer02}. In our
notation,
$\tilde{h}=2|\Lambda_{ii}|/D_{ii}=|H_i|\sqrt{8/D_{ii}}$. Both of these
coefficients have identical qualitative meaning: their large values
indicate fast decay of the FTLE fluctuations with $n$.

\begin{table} 
  \caption{\label{tab:ftle_fluct}Closest to zero Lyapunov exponents
    and fluctuation characteristics of the corresponding FTLEs.}
  \begin{indented}
  \lineup
  \item[]
  \begin{tabular}{@{}*{9}{l}}
    \br
    $h$ & $\ell$ & $\Lambda_\ell$ & $\Lambda_{\ell+1}$ &
    $D_{\ell\ell}$ & $D_{\ell+1,\ell+1}$ & $H_\ell$ & $H_{\ell+1}$ &
    $K_\ell$ \\
    \mr
    0.8 & 1 & 0.690 & \-0.499   & 0.00398 & 0.00398 &
    7.73 & \-5.59     & $\sim\!\! 10^{-10}$ \\
    1.2 & 3 & 0.00168 & \-0.692 & 0.0455 & 2.05     &
    0.00558 & \-0.342 & 0.428 \\
    1.7 & 5 & 0.0897 & \-0.105  & 0.646 & 2.03      &
    0.0789 & \-0.0521 & 0.150 \\
    \br
  \end{tabular}
\end{indented}
\end{table}

\begin{figure}
  \onefig{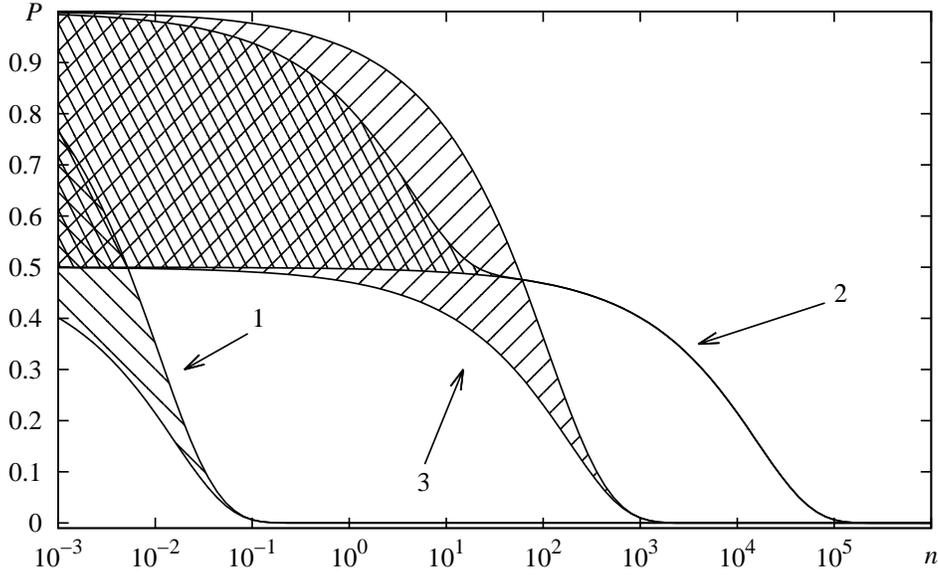}
  \caption{\label{fig:ftle_fluct} Lower and upper estimations of
    probability of wrong signs of FTLEs vs. the averaging interval
    $n$. Curves 1: hyperbolic chaos at $h=0.8$; curves 2: UDV at
    $h=1.2$; curves 3: chaotic dynamics at $h=1.7$.}
\end{figure}

Table~\ref{tab:ftle_fluct} collects the closest to zero Lyapunov
exponents, corresponding diffusion coefficients, and rates of
probability convergence for three values of $h$. The first row at
$h=0.8$ corresponds to the full synchronization where the oscillations
are hyperbolic, the second row at $h=1.2$ represents the case of UDV,
and the third row at $h=1.7$ illustrates the situation beyond the UDV
where the chaos, as discussed in \sref{sec:hyp}, is
non-hyperbolic. Using these data we compute the bounding intervals for
unstable manifold dimension fluctuations probability as functions of
the averaging length $n$, see Fig.~\ref{fig:ftle_fluct}. The curves 1
represent the hyperbolic chaos. Observe that the boundaries limiting
possible values of the fluctuations probability decay to zero and
vanish at $n=0.1$. It means that even at $n=1$ the fluctuations of the
unstable manifold dimension are absent. Using the Gaussian
approximation we can not detect the strict vanishing of the
fluctuations. Nevertheless, the observed rapid decay of the
fluctuations probability agrees well with the fact that the system is
actually hyperbolic in this case and the fluctuations have to be
absent. The curve 3 corresponds to the ``common'' non-hyperbolic
chaos. The fluctuations of the unstable manifold dimension are
essential only if we consider FTLEs at small $n$. They decay fast and
almost disappear at moderately high $n$. When $n>10^3$ the Gaussian
approximation predicts negligibly small fluctuations. The UDV case is
represented by the curve 2. The upper boundary of the probability
depends on $L_4$ that is far from zero and thus does not contribute
much to the fluctuations. The fluctuations of $L_4$ practically do not
influence the unstable manifold dimension variations when $n>10^2$
where the upper and lower boundaries merge. To the right of this point
the unstable manifold dimension fluctuations are basically determined
by the crossing zero $L_3$. The fluctuations decay very slowly and
remain at an essential level up to $n>10^5$ that is two order higher
then in the ``common'' case. Notice that the corresponding probability
convergence rate $H_3$ is one order less than the smallest rate
corresponding to the other FTLEs.

Another, rather heuristic, estimate for the magnitude of the unstable
dimension fluctuations provide the diffusion coefficients
$K_\ell=D_{\ell\ell}+D_{\ell+1,\ell+1}-2D_{\ell,\ell+1}$ that
characterise the probability of the negative value of the difference
$L_\ell(n)-L_{\ell+1}(n)$. As argued in~\cite{StatMechLyap11}, if
$\Lambda_\ell>\Lambda_{\ell+1}$ and $K_\ell=0$ the FTLEs are always
ordered correctly, $L_\ell(n)>L_{\ell+1}(n)$. Though it does not imply
the absence of the unstable dimension fluctuations directly (since
$L_\ell(n)$ still can be negative), it anyway indicates the low level
of ``bad'' fluctuations of corresponding FTLEs. One can see from
table~\ref{tab:ftle_fluct} that for the hyperbolic chaos $K_1$, as
expected, is effectively zero. The ``common'' chaos is characterized
by the moderately small $K_5$, and the highest value of $K_3$
corresponds to the UDV case.

\subsection{Angles between tangent subspaces}\label{sec:hyp}

The tangent space splitting can be introduced using covariant Lyapunov
vectors $\clvecfwd_i(t)$~\cite{CLV2011}. Let $\spaceA_k$ be a subspace
spanned by the first $k$ CLVs,
$\spaceA_k=\spanspc\{\clvecfwd_i(t)|i=1,\ldots,k\}$, and let
$\spaceB_{k+1}$ be a subspace of the rest of them,
$\spaceB_{k+1}=\spanspc\{\clvecfwd_i(t)|i=k+1,\ldots,m\}$. If there
are $k_{\mathrm{e}}$ positive Lyapunov exponents and UDV does not take
place, non-vanishing angles between the expanding subspace
$\spaceA_{k_{\mathrm{e}}}$ and the contracting one
$\spaceB_{k_{\mathrm{e}}+1}$ indicate uniform hyperbolicity. If UDV
occurs, some of CLVs either shrink or grow in time, but, nevertheless,
the tangent space splitting still can be introduced at arbitrary site
$k$. Varying $k$ and checking the angles between corresponding
subspaces, we can study the structure of the tangent space in this
case.

The presence of UDV automatically breaks the conditions of uniform
hyperbolicity, since an invariant splitting of the tangent space into
uniformly expanding and contracting subspaces can not exist. But, in
principle, the UDV does not exclude a tangent space splitting into
subspaces with strictly \emph{different} rates of expansion or
contraction. An $m$-dimensional attractor with UDV can have
$k$-dimensional strictly expanding subspace, and $(m-k)$-dimensional
subspace whose vectors can either decay or grow but always strictly
slower than in the first subspace. In this case angles between vectors
from these subspaces will be strictly non-zero. This situation is
known without relations to UDV and is referred to as partial
hyperbolicity~\cite{Pesin}. Partial hyperbolicity does not
automatically implies the properties peculiar to uniform
hyperbolicity, like, for example, structural stability. However, this
is the case provided that some additional requirements are
fulfilled~\cite{Pesin}. Below we shall check the partial hyperbolicity
for our system.

To analyze the possible occurrence of the tangencies between subspaces
$\spaceA_k$ and $\spaceB_{k+1}$, we have to find two vectors from them
having the smallest angle. The angle between these vectors is called
the first principal angle~\cite{GolubLoan}:
\begin{equation}
  \alpha_k=\min\{\angle(v_\spaceA,v_\spaceB)| v_\spaceA\in
  \spaceA_k, v_\spaceB\in \spaceB_{k+1}\}
\end{equation}
The tangency corresponds to the vanish of this angle, $\alpha_k=0$.

The straightforward way of computing $\alpha_k$, as described
in~\cite{HyperSpace08}, includes computation of the whole set of
CLVs. But it is much more efficient to consider the angles between
$\spaceA_k$ and the subspace $\spaceB_{k+1}^\perp$ which is the
orthogonal complement to $\spaceB_{k+1}$.  One have to compute only
the first $k$ orthogonal backward and forward Lyapunov vectors and
compose the matrix of their scalar products $\matr P(1\mycolon
k,1\mycolon k)$. The singular values of this matrix are the cosines of
the principal angles between $\spaceA_k$ and $\spaceB_{k+1}^\perp$. If
the smallest one is zero, $\sigma_{\mathrm{min}}=0$, there is a pair
of orthogonal vectors from $\spaceA_k$ and $\spaceB_{k+1}^\perp$. In
turn, it means that one can find a collinear pair of vectors from
$\spaceA_k$ and $\spaceB_{k+1}$, i.e., the tangency occurs. Thus, the
smallest principal angle between subspaces $\spaceA_k$ and
$\spaceB_{k+1}$ can be computed as
\begin{equation}
  \label{eq:min_pan}
  \alpha_k=\pi/2-\arccos(\sigma_{\mathrm{min}}).
\end{equation}
The details of this method can be found in~\cite{FastHyp12}.

\begin{figure}
  \onefig{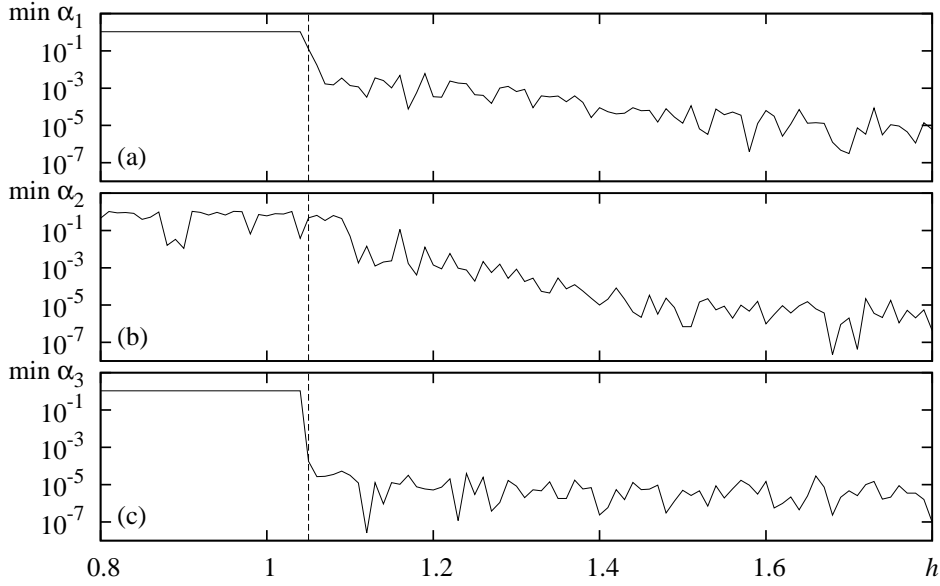}
  \caption{\label{fig:panh}The the smallest angles between tangent
    subspaces $\spaceA_k$ and $\spaceB_{k+1}$ found for $10^5$
    trajectory points. (a) $k=1$, (b) $k=2$, (c) $k=3$.  The vertical
    line marks the transition to the non-synchronous regime.}
\end{figure}

To detect the occurrences of the tangencies we fix a certain value of
the coupling parameter $h$, compute a sufficiently long trajectory of
the system~\eref{eq:chain}, find $\alpha_k$ at $t=nT$, which
corresponds to the consideration of the stroboscopic map, and take the
smallest value of the angle. In Fig.~\ref{fig:panh} this minimal angle
is plotted as a function of $h$ for $k=1$, $2$ and $3$. One can see
that in the synchronous regime all three first tangent directions are
well isolated from each other. The transition to the non-synchronous
regime is accompanied by the sharp drop of the all three angles. Their
behavior confirms the expected, due to UDV, violation of the uniform
hyperbolicity. Moreover, the drop of $\alpha_1$ indicates that the
subspaces $\spaceA_1$ and $\spaceB_2$ have tangencies. It means that
an invariant splitting with different expansion rates does not exist,
so that even partial hyperbolicity in a strict sense can not be
observed above the desynchronization threshold.

\begin{figure}
  a)\onefig{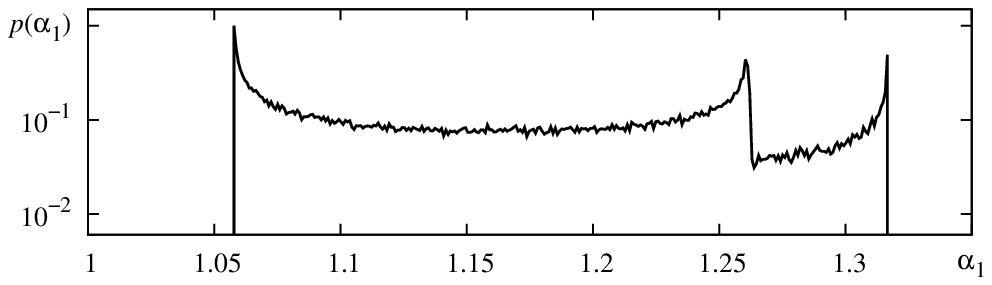}\\
  b)\onefig{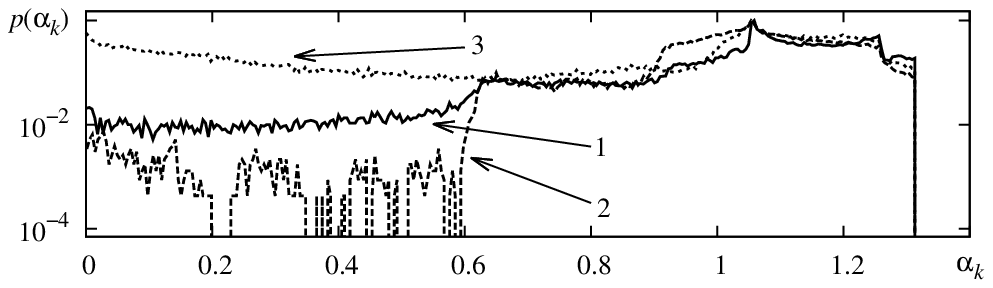}\\
  c)\onefig{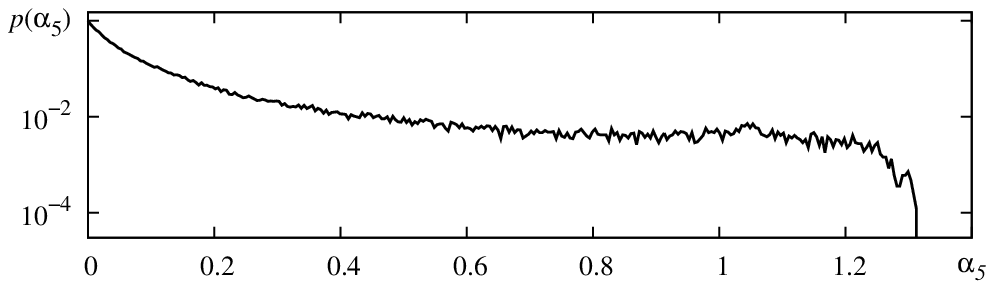}
  \caption{\label{fig:pan_dst}Distributions of the angle $\alpha_k$
    between subspaces $\spaceA_k$ and $\spaceB_{k+1}$ at (a) $h=0.8$,
    $k=1$, (b) $h=1.2$, $k=1,2,3$ (arrows 1, 2, 3 respectively), and
    (c) $h=1.7$, $k=5$. Each distribution is obtained for $10^5$
    trajectory points.}
\end{figure}

Distributions of $\alpha_k$ provide a more accurate information about
the angles between subspaces. If the distribution is separated well
from the origin, then a trajectory never approaches the trajectories
with exact tangencies, and one can conjecture that such ones are
absent at all. Figure~\ref{fig:pan_dst}(a) illustrates the case of the
synchronization at $h=0.8$, when $\spaceA_1$ and $\spaceB_2$ are the
expanding and contracting subspaces, respectively. The distribution of
the angle $\alpha_1$ between these subspaces is strictly isolated from
the origin that confirms the hyperbolicity of the synchronous
dynamics.

Figure~\ref{fig:pan_dst}(b) shows the case of UDV at $h=1.2$. The
distributions of the angles $\alpha_1$, $\alpha_2$, and $\alpha_3$
indicate the entanglement of the corresponding tangent subspaces. The
curve $p(\alpha_1)$ approaches zero being sufficiently small. It means
that though the trajectories with the exact tangencies between
$\spaceA_1$ and $\spaceB_2$ exist, they are very rare. The same is the
case for the subspaces $\spaceA_2$ and $\spaceB_3$. But $p(\alpha_3)$
is large at the origin. It means that the most of the tangencies
happens around the third covariant Lyapunov vector. Thus, one can
conclude that the chaos is non-hyperbolic in this case. However, the
rare vanishes of $\alpha_1$ and $\alpha_2$ indicate that the system
almost fulfills the partial hyperbolicity conditions.

Figure~\ref{fig:pan_dst}(c) illustrates the case $h=1.7$ where no
Lyapunov exponents fluctuate around zero and there are five unstable
directions. The expanding subspace is $\spaceA_5$ and the contracting
one is $\spaceB_6$. The distribution of the angles between these
subspaces $p(\alpha_5)$ is high at the origin. Hence, the attractor
contains a lot of trajectories with the exact tangencies between these
subspaces, i.e., it is non-hyperbolic.

\begin{figure}
  \onefig{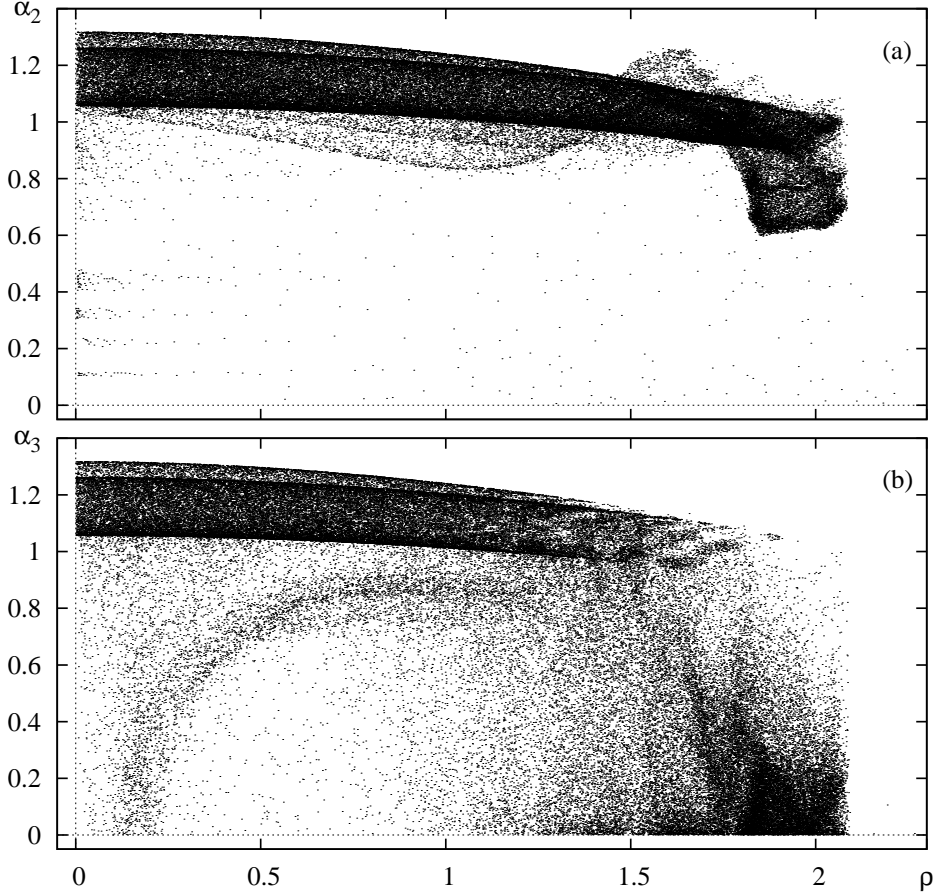}
  \caption{\label{fig:rhopan} Angle $\alpha_3$ and $\alpha_2$ vs.
    transverse distance to the synchronization manifold $\rho$
    computed for $10^5$ trajectory points. $h=1.2$}
\end{figure}

To reveal more information about the UDV case, in
Fig.~\ref{fig:rhopan} we plotted angles $\alpha_2$ and $\alpha_3$
vs. the distance to the synchronization manifold $\rho$ computed at
$10^5$ trajectory points. The major balks of points in both cases form
bold horizontal stripes. It means that the most probable angles are
located around 1 rad, and they can be encountered both close to the
synchronization manifold, and far from it. In the panel (a) beneath
the main stripe there is a series of sparse and hardly visible
horizontal stripes. We conjecture that they appear due to the passing
of the trajectory close to the embedded periodic orbits. In both
panels there are no dense cloud of points near the origin. Zero angles
mostly appear at a nonzero $\rho$. For example, in panel (b) one can
compare an arch-like structure approaching zero at $\rho\approx 0.2$
and a very dense collection of points around $\alpha_3=0$,
$\rho=2$. Hence, the tangencies basically occur when the system goes
far from the synchronization manifold. As one can see from
Fig.~\ref{fig:rholam}(b), since the third FTLE largely fluctuates in
this area of the phase space, a strong UDV takes place there. Thus the
tangencies in this area happen both at positive and at negative values
of this FTLE. Though seldom, $\alpha_3$ can nevertheless vanish near
the synchronization manifold, see Fig.~\ref{fig:rhopan}(b). In
Fig.~\ref{fig:rholam}(b) in this area the third FTLE can approach zero
from above but never become negative. Thus these vanishings of
$\alpha_3$ indicates the presence of homoclinic tangencies close to
the synchronization attractor.

\subsection{Structure of covariant Lyapunov vectors}\label{sec:clv}

The chaos in individual subsystems of the chain~\eref{eq:chain}
appears due to the uniform multiplication of phases, while amplitude
dynamics remains rather regular, see \sref{sec:local}. Thus, it is
natural for our system to compute CLVs with respect to the
perturbation of phases $\phi$ and amplitudes $r$. Given the Cartesian
coordinates $x$ and $y$, one can recompute a perturbation regarding
$\phi$ and $r$ as follows: $\tilde{r}=(\partial r/\partial
x)\tilde{x}+(\partial r/\partial y)\tilde{y}$,
$\tilde\phi=(\partial\phi/\partial x)\tilde{x}+(\partial\phi/\partial
y)\tilde{y}$,
\begin{equation}
  \label{eq:cart_polar}
  \tilde\phi=(-y\tilde x+x\tilde y)/r^2,\;\;
  \tilde r=(x\tilde x+y\tilde y)/r,
\end{equation}
where $r=\sqrt{x^2+y^2}$, and variables with tildes denote elements of
CLV, while those without tildes correspond to trajectory
coordinates. Thus, given the elements of CLV $\tilde a_n$ and $\tilde
b_n$, and corresponding trajectory coordinates $a_n$ and $b_n$, one
have to substitute each $\Re a_n$ and $\Im a_n$ instead of $x$ and
$y$, as well as $\Re \tilde a_n$ and $\Im \tilde a_n$ instead of
$\tilde x$ and $\tilde y$, and compute corresponding $\tilde \phi_n$,
$\tilde r_n$. The same have to be done for $b_n$.

\begin{figure}
  \onefig{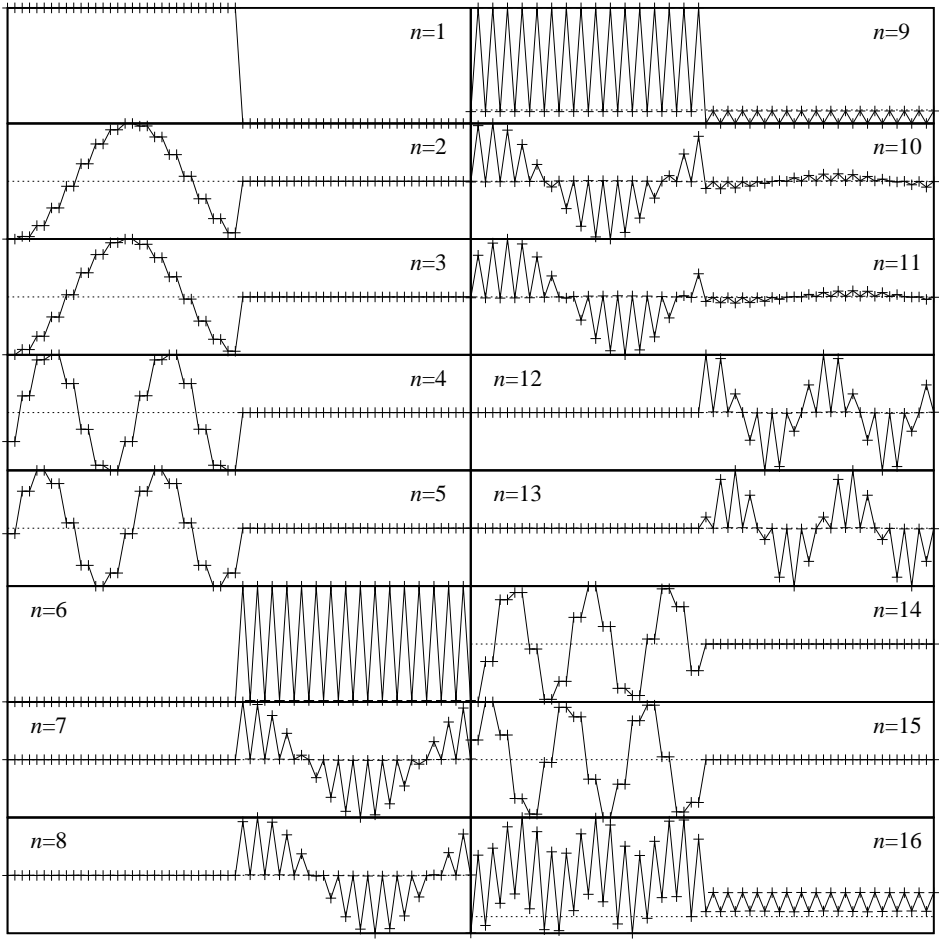}
  \caption{\label{fig:avgclv08} CLVs averaged over $10^5$ trajectory
    points at $h=0.8$. The vector numbers are shown in the panels. The
    system parameters are as in Fig.~\ref{fig:lyap}. In each panel the
    left 8 points correspond to perturbations of phase variables, and
    the right 8 ones represent amplitudes. The thin dashed lines show
    the zero axis}
\end{figure}

\begin{figure}
  \onefig{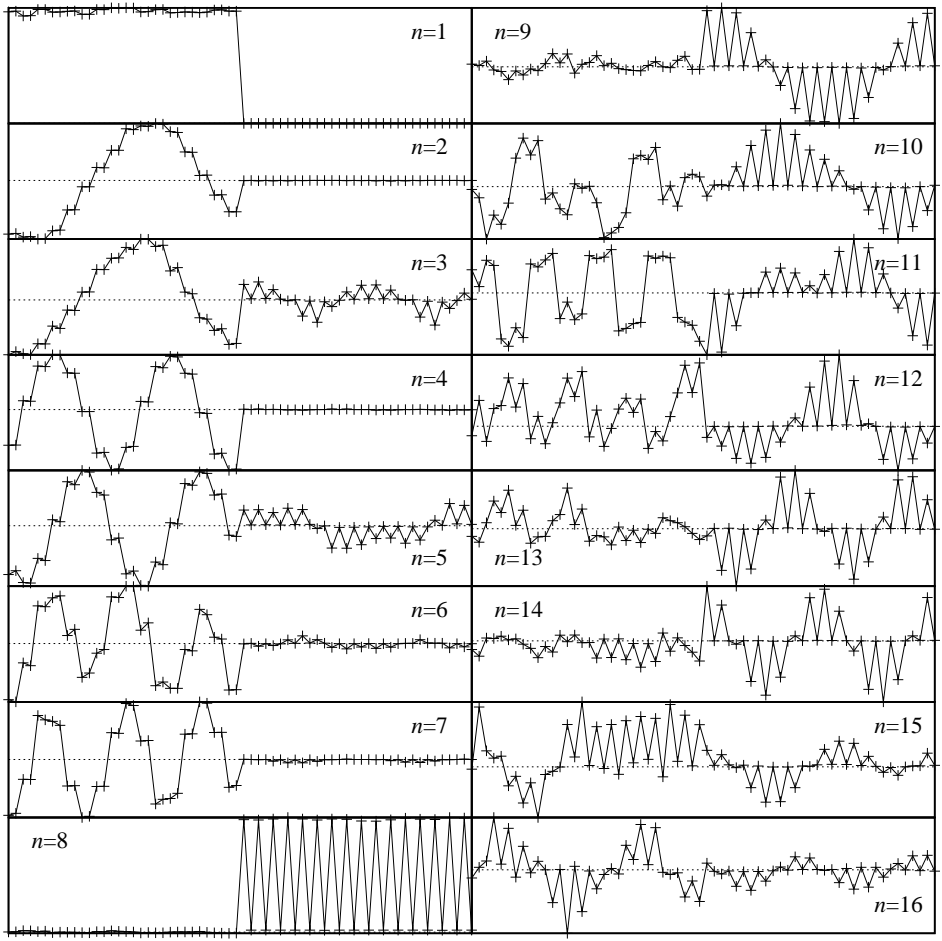}
  \caption{\label{fig:avgclv12} Averaged CLVs for $h=1.2$.}
\end{figure}

\begin{figure}
  \onefig{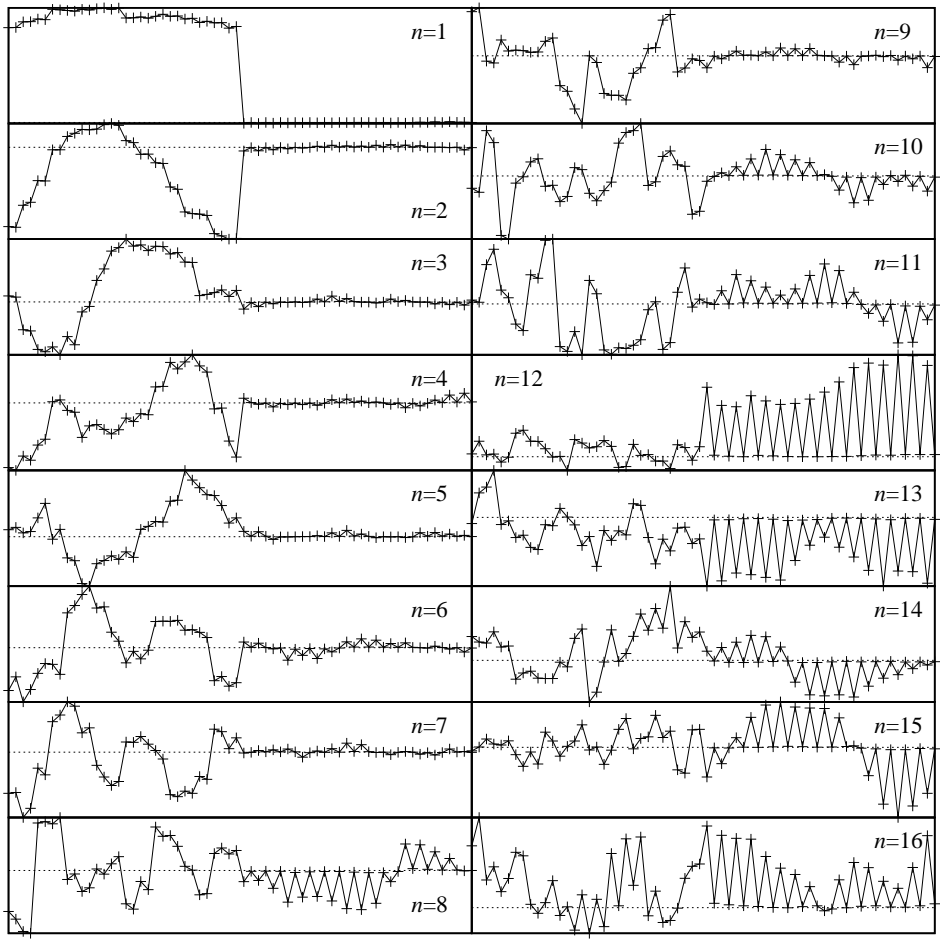}
  \caption{\label{fig:avgclv17} Averaged CLVs for $h=1.7$.}
\end{figure}

To visualize the structure of CLVs we find them for many trajectory
points, recompute with respect to phases and amplitudes, normalize and
average over the trajectory. The results are shown in
Figs.~\ref{fig:avgclv08}, \ref{fig:avgclv12}, \ref{fig:avgclv17}. When
chaos is hyperbolic and chain elements are synchronized, as in
Fig.~\ref{fig:avgclv08}, the CLVs are highly regular. The first one
corresponds to the single positive Lyapunov exponent. We observe that,
as expected, the expanding perturbation grows strictly along phase
variables. Notice high homogeneity of the vector. Phases in each site
of the vector grow identically which is a manifestation of the
synchronous regime. All other CLVs show the contracting
directions. Again the phase and amplitude perturbations are
isolated. The first four contracting vectors $n=2$,$3$,$4$, and $5$
represent the phase perturbation, the next three ones, $n=6$, $7$, and
$8$ are the amplitude ones, and so on. Vectors $n=9$, $10$, $11$, $16$
represent weakly mixed case: together with the phase perturbation
there is some perturbation along amplitude directions. But these
additions are small so that the phase directions dominate anyway.

The UDV case is illustrated in Fig.~\ref{fig:avgclv12}. One can see
that three first vectors are basically directed along the phase
subspace, i.e., the phases undergo the expansion. Observe high
similarity of the first five vectors with the previous case. After the
transition to the non-synchronous regime the structure of the
expanding vector $n=1$ remains almost unaltered, while the vectors $3$
and $5$ are rotated a little towards the amplitude directions. The
rotation of the third vector is in charge of the emergence of
entanglement of the expanding and contracting subspaces in this case
that is discussed in \sref{sec:hyp}. Two new phase perturbation
vectors $n=6$ and $7$ appear that are absent in the synchronization
regime in Fig.~\ref{fig:avgclv08}. The subsequent amplitude
perturbation vectors $n=8$, $9$, and $10$ correspond to the vectors
$n=6$, $7$, and $8$ in Fig.~\ref{fig:avgclv08}. Further correspondence
between CLVs in synchronous and UDV cases are rather untraceable by a
visual inspection.

Figure~\ref{fig:avgclv17} illustrates the ``common'' chaos observed
beyond hyperbolic and UDV cases. There are five positive Lyapunov
exponents and again, as before, all corresponding vectors are
basically represent the phase perturbation. The similarity of the
first seven vectors with the previous cases is still noticeable. The
first vectors consists of the homogeneous phase and zero amplitude
parts, i.e., the strongest expansion occurs along the diagonal in the
phase perturbation subspace. The vectors with indexes running from
$n=2$ to $7$ has more of less similar structure with the corresponding
vectors in the UDV case. Perhaps, the subsequent four vectors $n=8$,
$9$, $10$, and $11$ can be treated as new phase perturbation vectors,
while the vector $n=12$ corresponds to the vector $n=8$ in
Fig.~\ref{fig:avgclv12} and $n=6$ in Fig.~\ref{fig:avgclv08}.

\begin{figure}
  \onefig{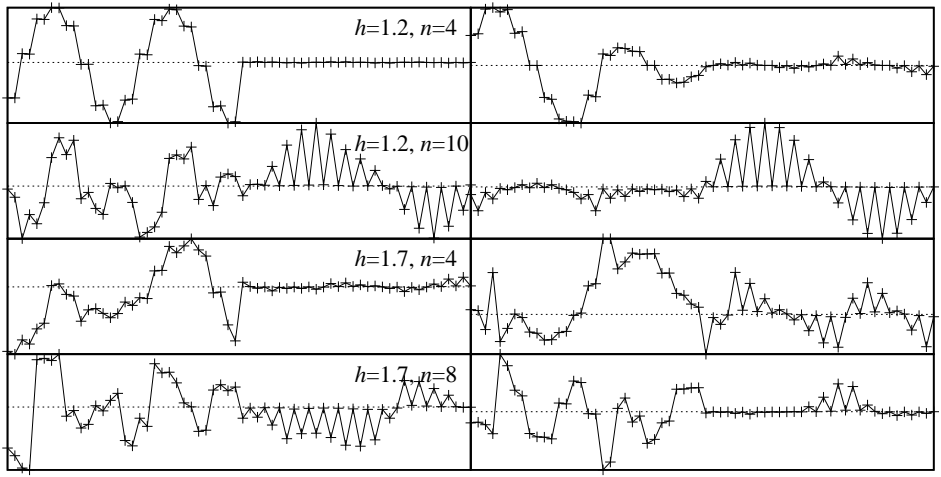}
  \caption{\label{fig:avgclv_tst} CLVs averaged over $10^5$ trajectory
    points (the left column of panels) and over $10$ points (the right
    column)}
\end{figure}

In principle, the isolation of phase and amplitude perturbations,
i.e., small amplitude or phase components of the averaged vectors,
could be obtained if their fluctuations along the trajectory were high
and uniformly distributed. But this is not the case for our
system. The fluctuations of the phase and amplitude components are
sufficiently small.  To test it, we have computed in the same way CLVs
averaged over just $10$ trajectory points. The vectors for synchronous
hyperbolic case at $h=0.8$ look almost indistinguishable from those
shown in Fig.~\ref{fig:avgclv08} up to a shift of maxima obviously
explained by the periodic boundaries conditions. For two other case
the coincidence is not so perfect but is still very
good. Figure~\ref{fig:avgclv_tst} shows the most different vectors:
the left columns reproduces the vectors from Figs.~\ref{fig:avgclv12}
and \ref{fig:avgclv17} and the right one shows the corresponding
vectors with the poor averaging. One sees that these vectors are
qualitatively similar.

\section{Summary and discussion}\label{sec:sum}

We discussed the violation of the synchronous uniformly hyperbolic
chaos in a chain of diffusively coupled hyperbolic chaotic
oscillators. The most important feature of these oscillators is that
they admit a physical implementation, so that our results can be
tested experimentally.

We observed the following scenario. The breakdown of the
synchronization begins when the second and the third Lyapunov
exponents simultaneously approach zero as the coupling strength
decays. The second FTLE starts to fluctuate, however the sign
changings are rare, and the range of coupling where the fluctuations
occur is very narrow. Within this range the on-off intermittency is
observed induced by UDV. In turn, the UDV inside this range arises due
to the invariant orbits embedded into the synchronization
attractor. Each of these orbits loses its transverse stability at
certain coupling strength, so that the attractor contains subsets with
various dimensions of unstable and stable manifolds. As the coupling
decays further, the second FTLE leaves the vicinity of zero, while the
third FTLE fluctuates changing its sign. This happens within a very
wide range of the coupling parameter. These fluctuations are related
to the UDV arising due to so called emergent set. This is the set of
invariant orbits with various dimensions of unstable and stable
manifolds. These orbits are located far from the synchronization
manifold and embedded into the non-synchronous attractor. The Gaussian
approximation of probability density of fluctuations of unstable
dimension was analyzed, and the slow decay of the probability,
compared to non-UDV cases, was revealed.

The test of angles between tangent subspaces confirms the expected
violation of the uniform hyperbolicity in the UDV case. Moreover, the
tests show even the absence of partial hyperbolicity, which is not
prohibited in UDV regime. However, the vanishings of angles
responsible for this occur very seldom, so that our system in the UDV
regime is almost partially hyperbolic. The partial hyperbolicity does
not automatically implies the properties peculiar for the uniform
hyperbolicity. Further study is required to understand how exactly
this almost partial hyperbolicity manifests itself.

Typically, CLVs strongly oscillate changing their directions from
point to point. However, for our system, being considered regarded to
phases and amplitudes of partial oscillators, the CLVs remain almost
constant. Thus, we are able to discuss the transformation of their
structure with the decay of the coupling strength. The visual
inspection reveals the isolation of phase and amplitude
perturbations. This basically occurs for first vectors, while the
vectors with high numbers contains both phase and amplitude
components.

\ack The research is supported by RFBR-DFG grant 11-02-91334.

\section*{References}

\bibliography{udvhyp2}

\end{document}